\documentstyle[preprint,aps,psfig]{revtex}

\begin{document}

\newcommand{\be}{\begin{equation}}
\newcommand{\ee}{\end{equation}}
\newcommand{\bea}{\begin{eqnarray}}
\newcommand{\eea}{\end{eqnarray}}
\newcommand{\ba}{\begin{array}}
\newcommand{\ea}{\end{array}}
\newcommand{\sprime}{^\prime}
\newcommand{\dprime}{^{\prime\prime}}
\newcommand{\tprime}{^{\prime\prime\prime}}

\tighten
\preprint{\vbox{\hbox{IFP-777-UNC}
\hbox{hep-ph/9910522} 
\hbox{\today}
}
}

\title{Nonabelian Discrete Symmetries, Fermion Mass Textures
and Large Neutrino Mixing}
\author{Paul H. Frampton and Andrija Ra\v{s}in}

\address{{\it Department of Physics and Astronomy}}

\address{{\it University of North Carolina, Chapel Hill, NC 27599-3255}}

\maketitle

\begin{abstract}

Nonabelian discrete groups are an attractive tool to describe
fermion masses and mixings. They have nonsinglet representations which
seem particularly suitable for distinguishing the lighter generations
from the heavier ones. Also, they do not suffer from the
extra constraints a continuous group must obey, {\it e.g.} limits
on extra particles. Some of the simplest groups are the
nonabelian discrete subgroups of SO(3) and SU(2), the so called
dihedral groups $D_n$ and dicyclic groups $Q_{2n}$, which both
have only singlet and doublet representations. 
After studying which vacuum expectation value (VEV) directions of
representations of dihedral and dicyclic groups preserve which subgroups,
we construct a simple model
based on the group  $Q_6 \times Q_6$. The model reproduces the masses and
mixings of all quarks and leptons, including neutrinos.
It has a large mixing angle in the  $\mu - \tau$ neutrino sector, 
in accordance with the recent SuperKamiokande results, while
keeping a small quark mixing in the bottom - charm sector. The reason is
similar to the one found in the literature based on the SU(5)
group: the large {\it left} handed mixing angle in the lepton
sector corresponds to the large unphysical {\it right} handed
in the down quark sector. The large mixing is also responsible
for the different hierarchies of the two heaviest families
in the up and down sector, and can be summarized as the order
of magnitude relation:
$
{m_s \over m_b} \sim \tan(\theta_{\mu\tau}) \sqrt{{m_c \over m_t}} \,\, .
$

\end{abstract} 

\newpage

\section{Introduction}

The question of the origin of fermion masses and mixings is one
of the most pressing questions in the Standard Model. Namely,
fermion masses and mixings
are merely input parameters and in order to get a handle on so
many arbitrary parameters one has to go beyond the Standard Model.
One of the most promising ways to go there is to utilize flavor symmetries
so that one understands the mass couplings as
parameters of flavor symmetry breaking. The major hope of such
approach is to minimize the number of input parameters and therefore
have verifiable predictions. Such approaches have been pursued for
many years with the quark masses and mixings and sometimes with
charged lepton masses as well.

On the other hand, during the last couple of years we have seen a steady
increase
in quantity and improvement in quality of neutrino data, and moreover
the first compelling evidence that neutrinos do have a mass. The
SuperKamiokande atmospheric
data\cite{superkamiokande}  strongly indicate that there is a large
nonvanishing mixing of the 
muon neutrino. The same experiment and several others\cite{othersolar} 
also indicate presence of neutrino mixing of the electron neutrinos coming
from  the Sun, and there is further\cite{lsnd}, though not nearly as
strong\cite{carmen}, evidence of other
neutrino mixing phenomena.

The simplest explanation of such data is that neutrinos have a
nonvanishing mass,
and similarly to quarks, mix with other neutrinos
due to the misalignment of flavor and mass
eigenstates. The smallness of the overall
scale of neutrino mass compared to other observed fermions, can
be understood by the see-saw mechanism\cite{see-saw}. On the other
hand, the neutrino mixing is generally explained quantitatively with
different
neutrino mass textures and many such examples exist in the literature
(for recent reviews see\cite{smirnov,altarelliferuglio}). 

It is thus an important question how to address  all
fermion masses and mixings, including neutrinos, in a viable
and simple flavor theory.
One of the interesting questions that arises is how to explain the 
large mixing in the neutrino sector between the second and
third generation\footnote{We do not consider the possibility
of more than three neutrino species.}, when the corresponding mixing
angle in the quark sector is very small. However, the key is to observe
that the
observed quark mixing angles pertain to the left handed sector. It
might be that their right handed mixing angles are large but they
are unphysical and hence unobservable. However, in a larger, possibly
unified theory, the right handed quark sector might be related to the
left handed lepton sector. Such is the case in SU(5) for example, where
the ${\bf 5}$ representation contains a {\it right} handed quark
and a {\it left} handed lepton. An interesting set of textures exploring
this left-right relation in SU(5) is found in \cite{albrightbabubarr},
and still other may be found in \cite{su5other}. A particularly
interesting aspect of this approach is that the large mixing can come
solely from the neutrino Dirac mass, regardless of the details of
the right handed Majorana mass matrix. Namely, if in the basis where
the charged lepton mass matrix is diagonal,
the neutrino Dirac matrix
is of the form (with left handed lepton doublets multiplying from 
the left)

\be
m^\nu = \left(
\ba{ccc}
0 & 0 & 0\\
0 & 0 & \sigma\\
0 & 0 & 1
\ea
\right)
\ee
where $\sigma$ is a number of order unity, and zeroes represent the much
smaller neglected entries, the resulting light neutrino mass matrix
\be
m^\nu_{light} = m^\nu {\bf M}^{-1}_N m^{\nu T} = {1/M_{33}}
\left(
\ba{ccc}
0 & 0 & 0\\
0 & \sigma^2 & \sigma\\
0 & \sigma & 1
\ea
\right)
\ee
has only one nonvanishing eigenvalue, and a large mixing angle. 
Note that the result does not depend on details of the
right handed Majorana mass matrix $M$, so long as $M_{33} \neq 0$.

Such models must in general use, 
additional global symmetries in order to get
the
desired texture. The question then arises if one could use only 
{\it discrete}
symmetries\cite{thomaswood}
 and build such textures. Such symmetries may be the
only remnant of the grand unified or stringy origin of an underlying
theory. Especially suitable are non-abelian
discrete symmetries, which in general prove to be more restrictive and
thus
more predictive. Another advantage of discrete symmetries is that one can
explain their origin as remnants of a broken gauge symmetry, or more
generally, of a string theory. The permutation symmetry group $S_3$
has been often used for
building neutrino mass matrices\cite{s3}.
Other models based on
nonabelian discrete groups have, surprisingly, not been
studied much in connection with quark mass textures
\cite{quarknonabelian,framptonkephart,framptonkong,caronelebed} and even
less so for 
neutrino mass textures: the quaternionic group $Q$ and dicyclic $Q_6$ 
has been studied in
\cite{changsenjanovic} in connection with the neutrino magnetic moment, 
and $\Delta(75)$, a discrete subgroup of SU(3)
in \cite{schmaltz}.

In this paper we study flavor theories with nonabelian discrete subgroups
of a gauged SU(2) in order to obtain suitable textures for both quark and
lepton mass matrices. Such groups have other motivations as well:
cosmological applications, i.e. such as Alice strings\cite{schwarz};
also the SU(2) origin might be interesting alternative starting point
for grand unification. 

In Section 2 we review and compare two
sets of discrete groups: the dihedral groups $D_m$ which
are subgroups of SO(3), and dicyclic groups $Q_{2n}$ which are subgroups
of SU(2). As we will see, just as the SU(2) is the spinorial 
generalization of SO(3), so are the $Q_{2n}$ spinorial generalizations
of $D_{2n}$, and we argue why these groups should be
taken seriously in model building. 

There are two reasons why $D_{m}$ and $Q_{2n}$ groups are 
particularly interesting for building flavor models. One is the
presence of only singlet and doublet representations. This
enables one to accommodate the three generations minimally in
such representations, and at the same time somehow distinguish
the heavy third family from the other two\cite{framptonkephart}. 
The second reason is
that these groups, even for small $m$ or $n$, have a rich structure
of subgroups, making it possible to use the scales of symmetry
breakings as the origin of fermion mass and mixing hierarchies.

In Section 3 we enumerate the possible
symmetry breaking directions of the vacuum expectation values (VEVs).
In Section 4 we present a model based on $Q_6 \times Q_6$ 
for which in Section
5 we demonstrate desirable quark mass and mixing hierarchies and
an acceptable large neutrino mixing, in a similar way as in
\cite{albrightbabubarr}.
Finally, Section 6 is the conclusion.

\bigskip
\bigskip

\section{Nonabelian discrete subgroups of SO(3) and SU(2)}

\underline{Dihedral} groups $D_m$ are the groups of rotations of a regular
$n$-agon in three dimensions (regarded as a two-faced entity - 
``dihedral") and are discrete nonabelian subgroups of
SO(3). They can be defined as

\be
D_m = \{ a, b | a^m = e, b^2 = e, aba = b \}.
\ee
They  are of order $2m$. Simplest examples are $D_2$ which has four
elements and is isomorphic to $K = Z_2 \times Z_2$ (the Klein group),
$D_3$ has six elements and is isomorphic to the permutation group $S_3$,
$D_4$ has eight elements and represents the symmetries of a square, etc.

Subgroups of $D_m$ include a $Z_m$ (generated by $a$) and $m$ $Z_2$'s 
(each one generated by $ba^l$, $l=0,1,...,m-1$) and further subgroups
depending on whether $m$ is prime or not. For example, $D_{2n}$ has in
addition two subgroups $D_n$ (one generated by $a^2$ and $b$, and the
other generated by $a^2$ and $ba$) and $n$ Klein (=$Z_2 \times Z_2$) 
groups (each
generated by $a^n$ and one of the $ba^l$, $l=0,...,n-1$), and
possible further subgroups if $n$ is not prime.

Particularly useful matrix representation for the groups $D_{2n}$
is given by
\be
a = 
\left(
\ba{ccc}
\cos\theta & -\sin\theta & 0\\
\sin\theta & \cos\theta & 0\\
0 & 0 & 1
\ea
\right)
\,\, , \,\,
b = \left( \ba{ccc} 1 & 0 & 0\\ 0 & -1 & 0\\
0 & 0 & -1
\ea
\right)
\label{o3matrix}
\ee
where $\theta = 2\pi/m$.

\underline{Dicyclic} groups $Q_{2n}$ are spinorial generalizations of
$D_{2n}$
and are subgroups of SU(2). They can be defined as
\be
Q_{2n} = \{ a, b | a^{2n} = e, b^2 = a^n, aba = b \}.
\ee
and are of order $4n$. The spinorial generalization may be seen
here from the property $b^4=e$ (compared to $b^2 =e$ in $D_m$). 
Smallest dicyclic groups include $Q_2$ which has four elements and
is isomorphic to $Z_4$, $Q_4$ with eight elements and isomorphic
to the quaternionic group, etc.

Subgroups of $Q_{2n}$ are $Z_{2n}$ (generated by $a$)
and $n$ $Z_4$'s (each generated by $a^n$ and one of $ba^l$, $l=0,...,n-1$)
and more subgroups depending on whether $n$ is prime or not.

For $Q_{2n}$
the matrix representation (\ref{o3matrix})
generalizes to 
\be
a =
\left(
\ba{cc}
e^{i\theta/2} & 0 \\
0 & e^{-i\theta/2}
\ea
\right)
\,\, , \,\,
b = \left(
\ba{cc}
0 & 1 \\
-1 & 0
\ea
\right)
\label{pin2matrix}
\ee
where $\theta = 4\pi/2n$.

Representations of both $D_{2n}$ and $Q_{2n}$ have four singlets and
$n-1$ doublets,
while representations of  $D_{2n+1}$ have 2 singlets and $n$ doublets.
As we wish to compare the dihedral and dicyclic groups of the same
order, we will not consider $D_{2n+1}$ further. 
Let us denote the four singlets $1, 1\sprime, 1\dprime,
1\tprime$ and the doublets as $2_i$, $1=1,...,n-1$.
The multiplication table for the representations of both $D_m$ and
$Q_{2n}$ can be found in \cite{framptonkephart,kongdisertation}
The behavior of the odd-numbered doublets
({\it i.e.} $2_{2i+1}$) is spinorial, and in the special case when
n is odd,  $1\dprime$ and $1\tprime$ are spinorial too.

It is useful to exhibit the decomposition of the SO(3) and SU(2)
representations under the $D_{2n}$ and $Q_{2n}$. We start by
identifying the ${\bf 2}$ of SU(2) 
with the first spinorial doublet $2_1$ of $Q_{2n}$, and then
find the decomposition of other SU(2) irreps from 
the decomposition of products of ${\bf 2}$'s. For SO(3)
it is necessary to identify ${\bf 3}$  with $1\sprime+2_1$.
Notice that the $Q_{2n}$ matrix representation in 
(\ref{pin2matrix}) is the
2-dimensional representation $2_1$ of $Q_{2n}$,
while 3-dimensional representation (\ref{o3matrix}) of $D_{2n}$ 
contains representations $1\sprime + 2_1$
of $D_{2n}$.
The decomposition of the lowest SO(3) and SU(2)
representations under $D_{2n}$ and $Q_{2n}$ are as follows
\be
\ba{ccc|ccc}
	SO(3) & \rightarrow & D_{2n}  & SU(2)
		& \rightarrow& Q_{2n}  
\\
\hline
	& & &
		{\bf 2} & \rightarrow & 2_1
\\
	{\bf 3} & \rightarrow & 1\sprime+2_1 & 
		{\bf 3} & \rightarrow & 1\sprime+2_2
\\
	& & &
		{\bf 4} & \rightarrow & 2_1+2_3
\\
	{\bf 5} & \rightarrow & 1+2_1+2_2 & 
		{\bf 5} & \rightarrow & 1+2_2+2_4
\\
	& & &
		{\bf 6} & \rightarrow & 2_1+2_3+2_5
\\
	{\bf 7} & \rightarrow & 1\sprime+2_1+2_2+2_3 & 
		{\bf 7} & \rightarrow & 1+2_2+2_4+2_6
\\
	& & &
		{\bf 8} & \rightarrow & 2_1+2_3+2_5+2_7
\\
	{\bf 9} & \rightarrow & 1+2_1+2_2+2_3+2_4 & 
		{\bf 9} & \rightarrow & 1+2_2+2_4+2_6+2_8
\ea
\label{q2n}
\ee	
etc., for sufficiently high $n$. Of course, for a given $D_{2n}$ 
(or $Q_{2n}$), there are only $n-1$ doublets, and for the doublets
with indices higher than $n-1$ one has to
use the identifications
 $2_n \equiv 1\dprime+1\tprime$, $2_i \equiv 2_{2n-i}$
and $2_0 \equiv 1 + 1\sprime$.
The odd-dimensional SU(2) representations cannot contain
spinorial doublets of $Q_{2n}$, since they are
vectorial. On the other hand, in $D_{2n}$ there
is no notion of spinors and all doublets appear in the
representations of SO(3) which are all odd-dimensional.
Since the highest doublet representation
is $2_{n-1}$ (and first appears in ${\bf 2n-2}$ of SU(2)) the
decomposition along the $T_3$ direction
is actually mod($n$), which correspond to the subgroup $Z_{2n}$
for $Q_{2n}$ (because of the half-integer isospins in spinorial 
representations). Similarly, the decomposition under $T_3$ of SO(3)
also shows the decomposition of irreps under $D_{2n}$ as elements of
the $Z_{2n}$ subgroup (notice that here the highest doublet $2_{n-1}$
first appears in  ${\bf 4n-4}$ of SO(3)).

We see that from model building viewpoint both groups have their
advantages. $Q_{2n}$ groups offer
more choice in putting generations in complete $SU(2)$ 
multiplets\footnote{This flavor $SU(2)$ group is not to be confused
with the usual electroweak $SU(2)_W$.}
without going to higher representations, where one has to
worry about the constraints on extra matter or anomaly cancellations.
Namely, in order to cancel anomalies
choosing to put the generations of fermions in a complete SU(2)
representations will satisfy the cancellation conditions
linear in SU(2). Other anomaly cancellation 
conditions do not have to be necessarily satisfied 
in a low-energy effective theory because of the in principle
unknown contributions from heavy fermions\cite{banksdineibanez}.
Thus,
we can put the three generations into $1+1+1$ or $1+2$ or a $3$
of SU(2), while in SO(3) we use only $1+1+1$ or $3$.
On the other hand $D_{2n}$ groups have more subgroups, and
therefore more choice in symmetry breaking. 
It will then depend on the underlying theory to decide whether
to pick a $SU(2)$ or SO(3) group. 

In the next two sections we first list the symmetry breaking directions,
and then we build a model based on $Q_6\times Q_6$.

\section{Symmetry breaking}

An interesting thing to note in
(\ref{q2n}) is that
${\bf 5}$ of SO(3) (or SU(2)) contains a singlet under {\it any} $D_{2n}$
or $Q_{2n}$. Indeed, the VEV of the singlet in 
{\bf 5} written as  a symmetric traceless $3 \times 3$ matrix
\be 
<{\bf 5}> = v {\rm diag}(1,1,-2)
\ee
does leave invariant a larger group. This group is generated by
the matrix representation of $D_{2n}$
in (\ref{o3matrix}), with any $\theta$ from $0$ to $2\pi$.
This invariance group,
loosely speaking, consists of a $SO(2)$ , generated by 
by the matrices $a$ in Eq. (\ref{o3matrix})
and a $Z_2$, generated by $b$ in Eq. (\ref{o3matrix}).
These
do not commute. More precisely, in the case of SO(3)
the subgroup is O(2) with two connected components:
one consists of all rotations around the z-axis and
is connected to the identity of SO(3), and the other one has
$180^\circ$ rotation around an axis in the x-y plane.
This group is generated by elements from (\ref{o3matrix})
for any $\theta$ between $0$ and $2\pi$.
In the case of SU(2) this generalizes to the so-called Pin(2) 
group (generated by elements from (\ref{pin2matrix}) for any
$\theta$ between $0$ and $4\pi$.)
which has interesting applications for Alice strings in
cosmology\cite{schwarz}. 

Next, notice that for a given n, the first SU(2) representation which
contains a singlet of $Q_{2n}$ but not of $Q_{2n+2}$,$Q_{2n+4}$,etc., is
${\bf 2n+1}$, since the highest doublet in it, $2_{2n}$, is 
identified by $2_{2n} \equiv 2_0 \equiv 1+1\sprime$, and similarly 
${\bf 4n+2}$ of SO(3) for $D_{2n}$.
Thus
\be
SO(3) \stackrel {\bf <4n+2>} {\longrightarrow} D_{2n} \,\, , \,\,
SU(2) \stackrel {\bf <2n+1>} {\longrightarrow} Q_{2n} \,\, .
\ee

We next identify the symmetry
breaking directions within $D_{2n}$ and $Q_{2n}$.

\vspace{1.0cm}

\noindent
{\bf Symmetry breaking in $D_{2n}$}

The $1\sprime$ is part of the triplet in SO(3).
The triplet is
\be
3 = 
\left(
\ba{c}
2_1 \\
1\sprime
\ea
\right)
\ee
and as can be seen from (\ref{o3matrix}), a VEV of $1\sprime$ 
preserves all transformations involving $a = R_{12}(2\pi/2n)$
(but not the ones involving $b$) and therefore preserves
$Z_{2n}$. On the other hand $2_1$ preserves one of the $2n$ $Z_2$
subgroups generated by $ba^l$ ($l=0,1,...,2n-1$) depending
on the VEV direction. In particular
$<2_1> = (v,{{cos(2\pi l/2n)-1}\over{sin(2\pi l/2n)}}v)^T$ preserves
the $Z_2$ generated by $ba^l$ ( $(ba^l)^2 = 1$ for any $l=0,1,...,2n-1$).
Since this $Z_2$ is not invariant under the $1\sprime$, simultaneous
VEVs of $1\sprime$ and $2_1$ will break $D_{2n}$ completely.

The VEV of the next doublet $2_2$ conserves one of the $n$ Klein
(=$Z_2 \times Z_2$) 
subgroups  generated by $a^n$ and $ba^l$
since $(a^n)^2 = 1$ and $(ba^l)^2=1$). This can be 
seen from the
decomposition of ${\bf 5} = 1 + 2_1 + 2_2$ under $D_{2n}$
\be
{\bf 5} = 
\left(
\ba{ccc}
v & 0 & 0 \\
0 & v & 0 \\
0 & 0 & -2v
\ea
\right)
+
\left(
\ba{ccc}
0 & 0 & v_1 \\
0 & 0 & v_2 \\
v_1 & v_2 & 0
\ea
\right)
+
\left(
\ba{ccc}
w_1 & w_2 & 0 \\
w_2 & -w_1 & 0 \\
0 & 0 & 0
\ea
\right)
\ee
The VEVs of the singlet and the doublet $2_1$ were discussed above. 
The last term is the $2_2$ of $D_{2n}$. Its VEV is invariant
under a Klein group generated by
$a^n = {\rm diag}(-1,-1,1)$ and $ba^l$ ($l=1,2,...,n-1$),
provided 
$w_2 = {{cos(4\pi l/2n) - 1} \over {sin(4\pi l/2n)}} w_1$.
When there is no relation between $w_1$ and $w_2$, the VEV of
$2_2$ simply leaves invariant a $Z_2$ subgroup generated by $a^n$.

So far we have shown which symmetry breaking directions are
possible with $1\sprime$, and the two doublets $2_1$ and $2_2$.
In the same manner one can analyze other doublets $2_i$, with $i=
3,4,...,n-1$, however as we will concentrate in the next section
on $Q_6$ and $D_6$ which only have $2_1$ and $2_2$, we do not study
further doublet VEVs here. 

The only thing that remains is to show 
which symmetries remain unbroken when $1\dprime$ or $1\tprime$
get a VEV. We have seen before that in
multiplication tables for representations $1\dprime+1\tprime$
behave as the (fictitious) doublet $2_n$ of $D_{2n}$ or $Q_{2n}$.
Also, as we have seen before, $2_n$ is in the ${\bf 2n+1}$
representation of SO(3) and corresponds to $T_3 = \pm n$.
This tells us that they preserve the symmetry
$Z_n$ (generated by $a^2$)   of $D_{2n}$.
In addition, when the VEVs of both $1\dprime$ and $1\tprime$
are equal, in $D_{2n}$ they will leave invariant also $b$, so
that the total symmetry left invariant by VEVs of $1\dprime+1\tprime$
in $D_{2n}$ is $D_n$.

We can summarize the symmetry breaking chains as follows
\be
\ba{ccccc}
	D_{2n} & \stackrel {< 1\sprime > } {\longrightarrow} 
		&
		Z_{2n}=\{ a \} \,\, ,&&
\\
	& & & &
\\
	D_{2n} & \stackrel 
		{ < 2_1 > = 
		\left(
		\ba{c}
		v \\
		{{c_l -1} \over {s_l}}v
		\ea
		\right) } {\longrightarrow} 
		& Z_{2}= \{ ba^l \} \, , &  l=0,1,..,2n-1 \,\, ,&
\\
	& & & &
\\
	D_{2n} & \stackrel 
		{ < 2_2 > = 
		\left(
		\ba{c}
		v \\
		{{c_{2l} -1} \over {s_{2l}}}v
		\ea
		\right) } {\longrightarrow} 
		& K= Z_2 \times Z_2 = \{ a^n,ba^l \} \,\, , &
l=0,1,...,n-1&
\\
	& & & &
\\
	D_{2n} &
	\stackrel 
		{ < 1\dprime>=<1\tprime >}
		{\longrightarrow} & D_n= \{ a^2,b \}  &
	\stackrel 
		\,\, ; \,\, { < 1\dprime>\neq<1\tprime >}
		{\longrightarrow} & Z_n= \{ a^2 \}  
\ea
\label{d2nvevs}
\ee
where $c_l = \cos{2\pi l/2n}$ and $s_l = \sin{2\pi l/2n}$. 

\vspace{1.0cm}

\noindent
{\bf Symmetry breaking in $Q_{2n}$}

In SU(2), the triplet can be represented as $2\times2$
traceless matrix $\Delta$ that transforms as $\Delta \rightarrow U \Delta
U^\dagger$. The components of this triplet are
\be
1\sprime =
\left(
\ba{cc}
f_1 & 0 \\
0 & -f_1
\ea
\right)
\,\, , \,\,
2_2 = 
\left(
\ba{cc}
0 & f_2 \\
g_2 & 0
\ea
\right)
\ee
so that $< 1\sprime > = {\rm diag}(v,-v)$ is invariant under $a$
from (\ref{pin2matrix})  ($<1\sprime> \rightarrow a <1\sprime>
a^\dagger$), but not $b$, and therefore also preserves
$Z_{2n}$ This is also easily understood, since $1\sprime$ has a vanishing
third component of isospin, and the multiplication tables
allow for isospin multiplications up to mod(n).
On the other hand, the vectorial doublet $2_2$ of $Q_{2n}$ does
preserve one of the $n$ $Z_4$ symmetries of $Q_{2n}$ generated
by $ba^l$, $l = 0,1,..., n-1$ (since $(ba^l)^4 = (a^n)^2 = 1$ ), and this
is achieved when the VEV points in the direction
\be
<2_2> = 
\left(
\ba{cc}
0 & v \\
-v e^{4\pi i l/2n} & 0
\ea
\right)
\ee

The $2_1$ spinorial doublet of $Q_{2n}$
does not preserve any subgroup of $Q_{2n}$ so its non-zero VEV breaks
the group to nothing. However, one has to stress that
spinorial couplings itself often preserve further accidental
discrete symmetries, and we will make use of those in the next section.

Similarly to $D_{2n}$, $1\dprime$ and $1\tprime$ break $Q_{2n}$ 
to $Z_n$ generated by $a^2$, but cannot be left invariant under $b$.
To summarize,
\be
\ba{ccccc}
	Q_{2n} & \stackrel {< 1\sprime > } {\longrightarrow} 
		&
		Z_{2n}=<a> \,\, ,&&
\\
	& & & &
\\
	Q_{2n} & \stackrel 
		 {< 2_1 >} {\longrightarrow} 
		& nothing \,\, ,&&
\\
	& & & &
\\
	Q_{2n} & \stackrel 
		{ < 2_2 > = 
		\left(
		\ba{c}
		v \\
		- v e^{i4\pi l/2n}
		\ea
		\right) } {\longrightarrow} 
		& Z_4=<a^n,ba^l> &\, , l=0,1,...,n-1 \,\, ,&
\\
	& & & &
\\
	Q_{2n} &
	\stackrel 
		{ < 1\dprime>,<1\tprime >}
		{\longrightarrow} & Z_n=<a^2>  && \,\, .
\ea
\label{q2nvevs}
\ee

\section{The Model}

Several flavor models based on groups $Q_{2n}$
and $D_{2n}$ exist. Flavor theories based on
$Q_6$ and higher $Q_{2n}$ groups were
studied in \cite{framptonkephart,framptonkong}.
The quaternionic group and $Q_6$ were also used to study
neutrino magnetic moments\cite{changsenjanovic}. 

In building a model, we must be guided by simplicity,
but at the same time by predictivity as well. Building
models on a single low order flavor group, such as
$Q_6$ or $D_6$ will inevitably require some fine tuning
to explain all patterns of fermion masses and mixings,
simply because the number of parameters is too small.
On the other hand, using a too large group leads to
many free parameters which defeats the purpose. A
model which lies somewhere between these two extremes 
based on $D_6 \times D_6$ was presented recently in
\cite{caronelebed}. It is one of the purposes of this 
paper to present an alternative model based on
group $Q_6 \times Q_6$. This model utilizes the
advantage of the $Q_{2n}$ groups over $D_{2n}$'s: 
there are more possibilities for dividing the three
generations among the irreps so that the anomaly
cancellation requirements are trivially satisfied.
We will also show that the smaller number of symmetry
breaking possibilities is enough to build a viable
model.

Let us first show why a single $Q_6$ is not sufficient,
unless a fine tuning is imposed. Assuming there are no
extra generations, we have three possibilites for
the assignment of fermions:
$1+1+1$, $1+2_1$ and $1\sprime+2_2$. The left handed
quark doublets $Q^i_L$ cannot be assigned to $1+1+1$ since
then, if there is no fine tuning, the diagonalization
of mass matrices will require large left handed mixings
and therefore lead to large nondiagonal entries in the
Cabibbo-Kobayashi-Maskawa (CKM) matrix. Suppose next that $Q^i_L$ are in
$1+2_1$. Then the
right handed up quarks $u^i_R$ cannot be in $1+2_1$
since then all the up quark masses will be of the
same order because the product $2_1 \times 2_1$ contains
a singlet. If $u^i_R$ is in $1\sprime+2_2$ then
masses of the up and charm quark transform as one
of the representations $1\dprime+1\tprime+2_1$.
However, this would predict that the up and charm
masses are equal or at least of the same order.
Namely the VEVs of $1\dprime$ or $1\tprime$ couple with 
equal strength to both up and charm, and similarly
for the VEV of a $2_2$ (unless one fine tunes one of
the components to have zero VEV). Thus we can assign
the up quarks only to $1+1+1$. However, the down quarks
then cannot be in $1+1+1$ since that would predict
similar ratios of masses in both up and down sectors
in the absence of fine tuning. Down quarks cannot be assigned
to $1\sprime+2_2$ either because of the same reason as
the up quarks cannot, as explained above. Finally, assignment of down
quarks to $1+2_1$ would give all the three down quarks the
mass of the same order.  A similar discussion can be given
for the last case when the $Q^i_L$ are assigned to $1\sprime+2_2$.

Therefore, we conclude that in the absence of fine tuning,
a single $Q_6$ group is not enough to explain the quark masses and mixings,
let alone the lepton sector. We are therefore led to a bigger
group, possibly a product of $Q_6$ with another group. As a simple 
example we look at the direct product of two $Q_6$ groups.

The flavor model is based on the group $Q_6 \times Q_6$
where the Standard Model fermions (and right handed neutrinos)
are assigned into the following
representations:
$Q^i_L  =  (1+2_1, 1)$, 
$u^i_R  =  (1, 1+2_1) $,
$d^i_R  =  (1, 1\sprime+2_2)$, 
$L^i_L  =  (1\sprime + 2_2, 1)$,
$e^i_R  =  (1, 1+2_1)$ and
$\nu^i_R  =  (1,1\sprime + 2_2)$.

There are no new fermions added to the SM field content, except for
the right handed neutrinos $\nu^c$, which we assume get a large mass.
We notice that, if the $Q_6 \times Q_6$ originated from  $SU(2) \times
SU(2)$, the above assignment satisfies Witten's requirement\cite{witten} 
of even number of doublets in each SU(2) (corresponding to
number of $2_1$s in each $Q_6$).

We assume the global symmetry $Q_6 \times Q_6$ (which is a subgroup 
of a gauged $SU(2) \times SU(2)$ symmetry broken at a high scale
by a pair of ${\bf 7}$'s) itself gets broken at some very high
scale by a set of scalar fields, {\it flavons}, possibly
through the Froggatt-Nielsen mechanism\cite{froggattnielsen}. 
The Standard Model Higgs is neutral under the $Q_6 \times Q_6$. 
We assume that all of the flavons with ``minimal" $Q_6$ numbers
exist: $(1\sprime,1)$,$(1,1\sprime)$,$(2_2,1)$,$(1,2_2)$,$(2_1,1)$
and $(1,2_1)$, but no combined representations such as
$(1\sprime,1\sprime)$.

We assume the following pattern of symmetry
breaking
\be
Q_6 \times Q_6 
\stackrel {< 1\sprime >} {\longrightarrow}
Z_6 \times Z_6
\stackrel {< 2_2 >} {\longrightarrow}
Z_2 \times Z_2
\stackrel {< 2_1 >} {\longrightarrow}
nothing
\ee

The important thing to notice is that although 
no remnant of the $Q_6$ is left unbroken
after $2_1$ gets a VEV,
the first generation fermions will still be massless
at the tree level.
This is because the Yukawa couplings coming from the
product of two spinorial representations carry an additional
$Z_2$ symmetry, so that only some linear combinations of
fermion fields get a mass.
This symmetry is of course broken in other sectors
of the theory, so that the first generation masses will
be generated through loop corrections. Thus
the first generation masses will in general be suppressed
by some small loop factor, and the precise predictions 
here will depend crucially on the flavon sector. In the next Section
we concentrate on the masses and mixings of the heaviest two 
generations while we
only make general comments about the first generation 
and leave more detailed analysis for a future publication.

\section{Fermion Masses and Mixings in the Model}

We assume that the flavon VEVs satisfy the following hierarchy:
\be
	(<1\sprime>\equiv v\sprime) 
\approx 
	(<2_2> \equiv (v\sprime \sigma, -v\sprime \sigma))  
\,\, >> \,\, 
	(<2_1> \equiv (\epsilon_1 M,\epsilon_2 M))
\ee
with 
$\sigma \sim 1, \epsilon_i \ll v\sprime / M \ll 1 $.
 
We will assume that the VEVs in both $Q_6$s are comparable for
corresponding fields ({\it i.e.} $(2_2,1) \approx (1,2_2)$). This
assumption is natural if one assumes a ``left-right" symmetric
theory ($Q_6 \times Q_6$ or the primordial 
$SU(2) \times SU(2)$)\footnote
{In the usual left-right symmetric theory the
parameters were chosen so that the left-right symmetry is broken
maximally\cite{mohasenj}. Here we assume that the parameters
are such that the VEVs do not break the symmetry between
the sectors, so that their VEVs are equal in both sectors.}.

In the limit $\epsilon_i = 0$ we get
\be
\ba{c}
m^u  \sim  \left(
	\ba{ccc}
		0 & 0 & 0
		\\
		0 & 0 & 0
		\\
		0 & 0 & 1 
	\ea
	\right) \,\, , \,\,
m^d  \sim  \left(
	\ba{ccc}
		0 & 0 & 0
		\\
		0 & 0 & 0
		\\
		k^d \sigma &  -k^d \sigma & 1	
	\ea
	\right) \,\, , \,\,
m^e  \sim  \left(
	\ba{ccc}
		0 & 0 & h^e\sigma \\
		0 & 0 & -h^e\sigma \\
		0 & 0 & 1	
	\ea
	\right)
\,\, , \\
m^\nu  \sim  \left(
	\ba{ccc}
		g^\nu \sigma^2 
			& -g^\nu \sigma^2 
			& h^\nu \sigma 
		\\
		-g^\nu \sigma^2 
			& g^\nu \sigma^2 
			& -h^\nu \sigma 
		\\
		k^\nu \sigma & 
			-k^\nu \sigma & 1 	
	\ea
	\right)  
\,\, , \,\,
M_{\bf N}  \sim  \left(
	\ba{ccc}
		0 & g_N & h_N \lambda 
		\\
		g_N & 0 & h_N \lambda 
		\\
		h_N \lambda & 
			h_N \lambda & 1 	
	\ea
	\right)  
\ea 
\ee 
where $\lambda \equiv \sigma v\sprime / M \ll 1$, and $g,h,k$ are some
numbers of order one.
In the quark sector at this level only the top and bottom get their
mass, while the CKM matrix is equal to the unit matrix. Notice the
large mixing of the second and third generations in the down quark
and charged lepton sectors, which is similar to SU(5) 
scenarios\cite{albrightbabubarr,rasin97}. 

The ratio
of the bottom to top mass is given by $m_b/m_t = ({ v\sprime \over M})
({ v \over \bar{v}})$, where $v$ and $\bar{v}$ are the VEVs of
the Standard Model Higgs doublets that couple to down and up
sector respectively. Since we do not know the ratio of the two electroweak
Higgs VEVs, we can conclude only that 
${v\sprime \over M}  =  1 \sim 1/60. $ 
Similarly in the charged lepton sector
only the tau lepton gets its mass at this level, and it is of
the order of the b quark mass.

The last two matrices are the Dirac neutrino and the right handed
neutrino mass matrices and we need to find the mass matrix of
the light neutrinos. 
The inverse of the
Majorana mass matrix can be computed exactly
\be
M^{-1}_{\bf N} \sim  \left(
	\ba{ccc}
		({{h_N \lambda} \over g_N})^2   & 
		{1 \over {fg_N}} + ({{h_N\lambda} \over g_N})^2  & 
		- {{h_N\lambda} \over g_N} 
		\\
		{1 \over {f g_N}} + ({{h_N \lambda} \over g_N})^2  & 
		({{h_N \lambda} \over g_N})^2   & 
		- {{ h_N \lambda} \over g_N} 
		\\
		- {{h_N \lambda} \over g_N}  &
		- {{h_N \lambda} \over g_N}  &
		1
	\ea
	\right)  
\ee
where $f \equiv 1 / (1 - 2 (h_N \lambda)^2 / g^2_N)$ and an overall factor
of $f$ 
has been absorbed in the overall mass scale.
Then we find that the mass matrix for
the light neutrino masses 
reduces to the following form
\be
m^\nu_{light} = - m^\nu {\bf M}^{-1}_N m^{\nu T} 
\sim \left(
	\ba{ccc}
		g \sigma^2 
			& -g \sigma^2 
			& h \sigma 
		\\
		-g \sigma^2 
			& g \sigma^2 
			& -h \sigma 
		\\
		h \sigma & 
			-h \sigma & 1 + k \sigma^2 	
	\ea
	\right)
\ee
where as before $g,h,k$ are some numbers of order one. Notice
that the orders of magnitude in the light neutrino matrix do not
depend on the details of the Majorana mass matrix, a feature
similar to those of abelian flavor theories with positive
charges\cite{rsilva93}. 

A $45^\circ$ rotation of the first two lefthanded
leptons brings us to the following basis for the charged
leptons and the light neutrinos
\be
m^e \sim \left(
	\ba{ccc}
		0 & 0 & 0 \\
		0 & 0 & -\sqrt{2} h^e\sigma \\
		0 & 0 & 1	
	\ea
	\right)
\,\, , \,\,
m^\nu_{light} \sim \left(
	\ba{ccc}
			0 & 0 & 0 
		\\
			0 & 
			2 g \sigma^2 & 
			-\sqrt{2}  h \sigma
		\\
			0 & 
			-\sqrt{2}  h \sigma & 
			1 + k \sigma^2 	
	\ea
	\right)
\ee
leading to two
neutrino masses of the same order and one zero mass neutrino.
These matrices are next diagonalized leading to the maximal mixing
in the $\nu_\mu$-$\nu_\tau$ sector
\be
\tan\theta_{\mu\tau} \sim \sigma.
\ee
This is consistent with the results of SuperKamiokande.

Second generation fermion masses and mixing angles will
be generated by factors of order $\epsilon$. 
When the VEV of $2_1$ is turned on (i.e. $\epsilon_i \neq 0$),
it 
produces the following order of magnitude entries for
the mass matrices:
\be
\ba{c}
m^u \sim \left(
	\ba{ccc}
		g^u\epsilon_2^2 & -g^u\epsilon_1\epsilon_2 & h^u\epsilon_2
		\\
		-g^u\epsilon_1\epsilon_2 & g^u\epsilon_1^2 & -h^u\epsilon_1
		\\
		k^u\epsilon_2 & -k^u\epsilon_1 & 1 
	\ea
	\right)
\,\, , \\
m^d \sim \left(
	\ba{ccc}
		g^d\sigma\epsilon_2 & -g^d\sigma\epsilon_2 & h^d\epsilon_2
		\\
		-g^d\sigma\epsilon_1 & g^d\sigma\epsilon_1 & -h^d\epsilon_1
		\\
		k^d\sigma & -k^d\sigma & 1	
	\ea
	\right)
\,\, , \\
m^e \sim \left(
	\ba{ccc}
		g^e\sigma\epsilon_2 & -g^e\sigma\epsilon_1 & h^e\sigma \\
		-g^e\sigma\epsilon_2 & g^e\sigma\epsilon_1 & -h^e\sigma \\
		k^e\epsilon_2 & -k^e\epsilon_1 & 1	
	\ea
	\right)
\ea
\ee
where $g^i,h^i,k^i$ are some unknown couplings of order 1. 

It is important to remember that all three matrices have still 
one zero eigenvalue because of
the accidental symmetry of the Yukawa couplings. However,
this symmetry is broken in other sectors of the theory,
and so there will be small loop corrections to the
Yukawa matrices eventually generating the lightest mass
eigenvalues. We will comment on this in a minute.

The masses and mixings are to the leading order 
\be
\ba{c}
	{m_\mu \over m_\tau} \approx 
		(g^e-h^ek^e) \sigma \sqrt{\epsilon_1^2 + \epsilon_2^2}
			\sim \sigma \epsilon 
\,\, , \,\,
\\
	{m_s \over m_b} \approx 
		(g^d-h^dk^d) \sigma \sqrt{\epsilon_1^2 + \epsilon_2^2}
			\sim \sigma \epsilon
\,\, , \,\,
\\
	{m_c \over m_t} \approx (g^u-h^uk^u) (\epsilon_1^2 + \epsilon_2^2) 
			\sim \sigma \epsilon^2
\,\, , \,\,
\\
	\theta_{\mu\tau} \sim O(1) 
\,\, , \,\,
	\theta_{cb} \sim \epsilon
\label{orderrelns} 
\ea
\ee

More quantitative relations are obtained if one makes further
assumptions that  the order
one constants $g^i,h^i,k^i$ perhaps
vary by a factor of two or so. Also, the VEVs in the two different
$Q_6$s do not have to be equal in both sectors. For example, 
if the VEVs in $<(2_1,1)>$ and $<(1,2_1)>$ differ by a factor
of 3 we get relations
\be
\ba{c}
	{m_\mu \over m_\tau} \sim \sigma\epsilon
\,\, , \,\,
	{m_s \over m_b} \sim {{\sigma\epsilon} \over 3} 
\,\, , \,\,
	{m_c \over m_t} \sim {{\epsilon^2} \over 3} 
\\
	\theta_{\mu\tau} \sim O(1) 
\,\, , \,\,
	\theta_{cb} \sim \epsilon/3 \approx \sqrt{{m_c \over {3 m_t}}} 
\label{orderrelnsprecise}
\ea
\ee
The relation between the masses of charged leptons and down quarks is
the usual Georgi-Jarlskog relation. However, notice the additional factor
of $\sqrt{3}$ in the $\theta_{cb}$ expression which makes it in better
agreement with data.
However, one has to stress again the uncertainty with
factors of order one at this level of predictivity.

Let us remark on the size of parameters
$\epsilon \sim <2_1>/M$, which characterizes the size of the VEVs of
the spinorial doublets $2_1$, and $v\sprime/M$, 
which characterizes the size of the VEVs of the singlets $1\sprime$
and doublets $2_2$. Relations (\ref{orderrelns}) ( or
(\ref{orderrelnsprecise}) ) tell us that $\epsilon$ is
of order $1/20$ or so, using our assumption $\sigma \sim O(1)$.
This then tells us that $v\sprime/M$ is somewhere between
$1$ and $1/20$, and does not have to be small. However,
since we want to control the size of the first generation
masses (see next paragraph), we do not want this number to
be too close to $1$.

Finally, let us comment on the first generation masses and mixings.
Such terms can be generated at the tree level only by higher
dimensional
terms and will be suppressed. Also, the first generation masses and
mixings are not strictly zero at the loop level either since the
flavor symmetries of $Q_6$ and its subgroups are completely broken
by the VEVs of the $2_1$ doublets. This will generate masses at the
loop level, which will however be suppressed by the loop factors.
This will be for example notable in the down and charged lepton sector 
where the VEVs of the components of the $2_2$ doublet will not be
in general exactly equal to the negative of each other. 
Thus, we expect all first generation fermion masses, including
light neutrinos, to be smaller
compared to second and third generation masses. We also expect suppressed
mixing angles involving first generation fermions, including neutrinos,
which seems to favor the small angle MSW solution. We leave 
more precise statements concerning the first generation for a future
publication.

\bigskip
\bigskip

\section{Conclusions}

Nonabelian discrete groups are an attractive tool to describe
fermion masses and mixings. They have nonsinglet representations which
seem particularly suitable for distinguishing the lighter generations
from the heavier ones. Also, they do not suffer from the
extra constraints a continuous group must obey, {\it e.g.} limits on extra
particles
(gauge or pseudogoldstone, depending on the continuous group being
local or global). One of the simplest groups are the
nonabelian discrete subgroups of SO(3) and SU(2), the so called
dihedral groups $D_n$ and dicyclic groups $Q_{2n}$, which both
have only singlet and doublet representations. Such groups have
also a rich structure of subgroups which makes it possible to
use the hierarchies in the symmetry breaking chain as the origin
of hierarchies in fermion masses and mixings. Equations
(\ref{d2nvevs}) and (\ref{q2nvevs}) summarize which VEVs
break dihedral and dicyclic groups to which subgroups.

As an example,  we constructed a simple model based on the group 
$Q_6 \times Q_6$.  The model reproduces the masses and
mixings of all quarks and leptons, including neutrinos.
It has a large mixing angle in the 
$\mu - \tau$ neutrino sector, while keeping a small quark
mixing in the bottom - charm sector. The reason is similar
to the one found in the literature based on the SU(5) 
group\cite{albrightbabubarr}: the large
{\it left} handed mixing angle in the lepton
sector corresponds to a the large unphysical {\it right} handed
in the down quark sector. The large mixing is also responsible
for the different hierarchies of the two heaviest families
in the up and down sector, and can be summarized as the
order of magnitude relation
\be
{m_s \over m_b} \sim \tan(\theta_{\mu\tau}) \sqrt{ {m_c \over m_t} }
\,\, .
\ee

\bigskip
\bigskip
\bigskip

{\bf Acknowledgments  } \hspace{0.5cm} 

This work was supported in part by the US Department of Energy
under the Grant No. DE-FG02-97ER-41036. We thank 
Thomas W. Kephart, Richard F. Lebed and Goran Senjanovi\'{c}
for discussions.

{\bf Note Added} Another paper using nonabelian discrete symmetry
to describe quark and lepton masses appeared recently, using
the double tetrahedral group\cite{lebednew}.

\end{document}